\documentclass[prb,twocolumn]{revtex4}
\usepackage{graphicx,xcolor}

\newcommand{\be}{\begin{equation}}

\newcommand{\ee}{\end{equation}}
\newcommand{\bea}{\begin{eqnarray}}
\newcommand{\eea}{\end{eqnarray}}
\newcommand{\bse}{\begin{subequations}}
\newcommand{\ese}{\end{subequations}}

\begin{document}

\def\K{{{K}}}
\def\Q{{{Q}}}
\def\Gbar{\bar{G}}
\def\k{{{k}}}
\def\q{{\bf{q}}}

\newcommand{\tk}{\ensuremath{\tilde{k}}}
\def\K{{\bf{K}}}

\title{Dual Fermion Dynamical Cluster Approach for Strongly Correlated Systems}

%\author{ LSU$^{1}$, CPHT$^{2}$, and G\"ottingen$^{3}$}
\author{ S.-X. Yang$^{1}$, H. Fotso$^{1}$, H. Hafermann$^{2}$, K.-M. Tam$^{1}$, J.\ Moreno$^{1}$, T. Pruschke$^{3}$, and M.\ Jarrell$^{1}$}
\affiliation{$^{1}$Department of Physics and Astronomy, Louisiana State University, Baton Rouge, Louisiana 70803, USA}
\affiliation{$^{2}$Centre de Physique Th\'eorique, \'Ecole Polytechnique, CNRS, 91128 Palaiseau Cedex, France}
\affiliation{$^{3}$Department of Physics, University of G\"ottingen, 37077 G\"ottingen, Germany}

\begin{abstract}

We have designed a new multi-scale approach for strongly correlated systems 
by combining the Dynamical Cluster Approximation (DCA)
and the recently introduced dual-fermion formalism. This approach employs an exact mapping from a 
real lattice to a DCA cluster of linear size $L_c$ embedded in a dual fermion lattice. Short-length-scale 
physics is addressed by the DCA cluster calculation, while longer-length-scale physics is addressed 
diagrammatically using dual fermions.  The bare and dressed dual fermionic 
Green functions scale as ${\cal{O}}(1/L_c)$, so perturbation theory on the dual lattice converges 
very quickly.  E.g., the dual Fermion self-energy calculated with simple second order perturbation 
theory is of order ${\cal{O}}(1/L_c^3)$, with third order and three body corrections down by an additional factor
of ${\cal{O}}(1/L_c)$.    
\end{abstract}

%\pacs{74.20.-z, 74.20.Fg, 74.25.Dw, 71.10.-w}
%74.20.-z 	Theories and models of superconducting state
%74.20.Fg 	BCS theory and its development 
%74.25.Dw 	Superconductivity phase diagrams 
%71.10.-w 	Theories and models of many-electron systems

\maketitle

%==========BODY OF PAPER =========================================

\section{Introduction}
Dynamical mean-field theory \cite{DMFA1, DMFA2,DMFA3} has been remarkably successful at capturing the physics 
of strongly correlated systems dominated by spatially local correlations.  
Successes include the description of the Mott transition in the Hubbard model, screening effects 
in the periodic Anderson model, as well as the description of correlation effects in realistic 
systems~\cite{DMFA_reviews1,DMFA_reviews2,DMFLDA}. % we need to cite a LDA=DMFT review

Since the introduction of the Dynamical Mean-Field Approximation (DMFA) there have been a number 
of attempts to develop formal extensions around the DMFA that incorporate non-local corrections.  
These include cluster extensions of the DMFA, such as the Dynamical Cluster Approximation 
(DCA)~\cite{DCA1,DCA2,DCA3}  or the Cellular Dynamical Mean-Field Theory (CDMFT)~\cite{CDMFT}, and multi-scale 
approximations where the DMFA or DCA vertices are used to parameterize two-particle field 
theories and longer ranged correlations can thus be captured~\cite{DVA,MSMB,Kusunose}.  One of the main limitations 
of these theories is that they converge slowly with the linear cluster size $L_c$, especially for 
the calculation of transition temperatures.  

The Dual Fermion formalism~\cite{dual_fermion1, dual_fermion2, dual_fermion3} is however, distinctly different from other cluster extensions of the 
DMFA.  In the dual fermion formalism, the lattice action is first mapped onto a dual fermion action where the interaction 
vertices are the n-body reducible vertices of the cluster.  This mapping is exact, so the dual fermion 
formalism provides a complete and exact formalism for the lattice problem.  Thus far, the dual fermion formalism 
has only been explored  using the DMFA or the CDMFT as cluster solvers~\cite{cluster_DF}.  However, the CDMFT has the 
disadvantage in this context that it violates translational invariance, so that the CDMFT vertices 
are rank-4 tensors in the spatial or momentum indices, which are too large to be stored 
and manipulated on most computers, especially for large clusters. 
Thus in this manuscript we propose the Dual Fermion Dynamical Cluster approach (DFDCA), 
within which the long-ranged correlations can be systematically incorporated 
through the dual fermion lattice calculation.
%In this manuscript we show that the dual fermion formalism may be used 
%to systematically incorporate long-ranged correlations into the DCA.
%In this manuscript we show that the DCA may be used as a DF cluster solver which preserves the exact mapping from the lattice to the DF cluster. 
Since the DCA preserves the translational invariance of the lattice system, the DCA two-body 
vertices are rank-3 tensors which, for modest cluster sizes, will fit in the memory of modern computers.  
Another difference,  which we will discuss in detail, is that the small parameter 
for the DFDCA is the dual fermion single-particle Green function, which scales  as $G_d\sim {\cal{O}}(1/L_c)$ with 
$L_c$ being the linear cluster size.   As a result, perturbation theory on the dual fermion 
lattice converges very quickly.  Simple second order perturbation theory on the dual fermion lattice already yields a
dual fermion self-energy of order ${\cal{O}}(1/L_c^3)$ with two-body and three-body corrections down by an additional factor
of ${\cal{O}}(1/L_c)$.  Higher order approximations are also possible, since the, e.g., three-body 
vertex corrections to the DFDCA self-energy are small, ${\cal{O}}(1/L_c^4)$.  Therefore, the resulting DFDCA 
formalism converges very quickly with increasing cluster size, with corrections to the self-energy no larger than 
${\cal{O}}(1/L_c^4)$.

\section{Mapping the DCA formalism to dual Fermions} 
\label{sec:DFDCA_derivation}
We will derive the DFDCA formalism with the example of the Hubbard model.  Its Hamiltonian is  
\begin{eqnarray}
\label{hubbard}
H&=&-\sum_{<i j>}t_{ij}(c_{i\sigma }^{\dag}c_{j\sigma }+{\rm{h.c.}}), \nonumber  \\
% +\epsilon\sum_{i\sigma}n_{i\sigma} 
&+& U\sum_{i}(n_{i\uparrow}-1/2)(n_{i\downarrow }-1/2) 
\end{eqnarray}
where $t_{ij}$ is the matrix of hopping integrals, 
$c_{i\sigma }^{(\dag)}$ is the annihilation (creation) operator for electrons on lattice 
site $i$ with spin $\sigma$, $n_{i\sigma }= c_{i\sigma }^{\dag}c_{i\sigma } $, and $U$ the intra-atomic repulsion.  

The DMFA, and its cluster extensions such as the DCA, are based upon the common idea of embedding a 
cluster in a lattice.  We assume that the cluster, of size $N_c=L_c^D$, dimensionality $D$, sites labeled 
by $I$ and wavevectors $K$,  is embedded in a large but finite-sized lattice of size $N$ with
sites $i$ and wavevectors $k$.  In the DCA, the reciprocal space of the lattice is divided into 
$N_c$ cells of identical geometry and linear size $\Delta k$.  The cell centers are labeled by $K$, 
and the points surrounding $K$ within the coarse-graining cell are labeled with $\tk$.  We will also
invoke a dual space lattice which is of the same size and geometry as the real lattice.

%\begin{itemize}
%\item DMFT:
%\end{itemize}

The action for this model is 
\begin{equation}
S[c^{*},c]=-\sum_{\omega,k,\sigma}c_{\omega,k,\sigma}^{*}[(i\omega+\mu)\mathbf{1}
-h_{k}]c_{\omega,k,\sigma}+\sum_{i}S_{loc}[c_{i}^{*},c_{i}],
\end{equation}
where $S_{loc}[c_{i}^{*},c_{i}]$ is the local part of the action including the Hubbard interaction term,
 $c_{i}^{*}$ and $c_{i}$ are now Grassmann numbers corresponding to creation and annihilation operators on the
lattice, $\mu$ the chemical potential, $h_{k}$ the lattice bare dispersion,
and $\omega=(2n+1)\pi T$ the Matsubara frequencies.
Decomposing the wavevector according to $k=K+\tk$, the lattice action becomes
\begin{eqnarray}
S[c^{*},c]=\sum_{i}S_{loc}[c_{i}^{*},c_{i}]
\;\;\;\;\;\;\;\;\;\;\;\;\;\;\;\;\;\;\;\;\;\;\;\;\; \;\;\;\;\;\;\;\;\;\;\;\;\;\;\;\;\;\;\;\;\; &&\nonumber \\
\;\;\;\;\;\;-\sum_{\omega,K,\tk,\sigma}c_{\omega,K+\tk,\sigma}^{*}
[(i\omega+\mu)\mathbf{1}-h_{K+\tk}]c_{\omega,K+\tk,\sigma}. &&
\end{eqnarray}
The goal is to express this action in terms of the DCA cluster problem
\cite{explanation}
%\footnote{Note that since the interaction is assumed to be local, it is unaffected 
%by coarse-graining. Non-local interactions however will be coarse-grained.
%}
\begin{eqnarray}
S_{\textrm{cluster}}[\overline{c}^{*},\overline{c}]
=\sum_{I} S_{loc}[\overline{c}_{I}^{*},\overline{c}_{I}]
\;\;\;\;\;\;\;\;\;\;\;\;\;\;\;\;\;\;\;\;\;\;\;\;\; \;\;\;\;\;\;\;\;\;\; &&\nonumber\\
\;\;\;
-\sum_{\omega,K,\sigma}\overline{c}_{\omega,K,\sigma}^{*}[(i\omega+\mu)\mathbf{1}
-\overline{h}_{K}-\Delta(K,i\omega)]\overline{c}_{\omega,K,\sigma},&&
\label{eqn:sclust}
\end{eqnarray}
where $\overline{c}_{I}^{*}$ and $\overline{c}_{I}$ are now Grassmann numbers corresponding to creation 
and annihilation operators on the DCA cluster, and $\Delta(K,i\omega)$ is the cluster hybridization function.  
To this end, we add and subtract the hybridization function and coarse-grained 
dispersion, i.e.,
\begin{eqnarray}
&&\sum_{\omega,K,\tk,\sigma} c_{\omega,K+\tk,\sigma}^{*}
[\overline{h}_{K}+\Delta(K,i\omega)]c_{\omega,K+\tk,\sigma} \nonumber\\
&=&\frac{N}{N_{c}}\sum_{\omega,K,\sigma}\overline{c}_{\omega,K,\sigma}^{*}
[\overline{h}_{K}+\Delta(K,i\omega)]\overline{c}_{\omega,K,\sigma},
\end{eqnarray}
where the last line follows from the DCA coarse-graining identity
\begin{equation}
\overline{c}_{\omega,K,\sigma}^{*}\overline{c}_{\omega,K,\sigma}
\equiv \frac{N_c}{N} \sum_{\tilde{k}}c_{\omega,K+\tilde{k},\sigma}^{*}c_{\omega,K+\tilde{k},\sigma}
\end{equation}
and the coarse-grained dispersion is given by
\begin{equation}
\overline{h}_{K} = \frac{N_c}{N}\sum_{\tk} h_{K+\tk}.
\end{equation}
The DCA coarse-graining identity preserves the Fermionic Lie algebra, despite the fact 
that it is not a canonical transformation,
\begin{equation}
\left\{ \overline{c}_{K,\sigma}^\dag , \overline{c}_{K',\sigma'}   \right\} = 
\frac{N_c}{N} \sum_{\tk} \left\{ {c}_{K+\tk,\sigma}^\dag , {c}_{K'+\tk,\sigma'}   \right\} = \delta_{K\sigma,K'\sigma'},
\end{equation}
where the last step follows since the coarse graining cells surrounding $K$ and $K'$ have the 
same geometry and contain the same number of states which, therefore, may be labeled with the 
same $\tk$. We then obtain
\begin{eqnarray}
S[c^{*},c]  = \sum_{i}S_{loc}[c_{i}^{*},c_{i}]
\;\;\;\;\;\;\;\;\;\;\;\;\;\;\;\;\;\;\;\;\;\;\;\;\;\;\;\;\;\;\;\;&&\nonumber\\
 -\sum_{\omega,K,\tk,\sigma}c_{\omega,K+\tk,\sigma}^{*}[(i\omega+\mu)\mathbf{1}-\overline{h}_{K}
-\Delta(K,i\omega)]c_{\omega,K+\tk,\sigma}& & \nonumber \\
-\sum_{\omega,k,\sigma}c_{\omega,k,\sigma}^{*}
[\Delta(M(k),i\omega)+\overline{h}_{M(k)}-h_{k}]c_{\omega,k,\sigma}.\;\;\;\;\;\;\;\;\;\;\;\;\; & & 
\end{eqnarray}
In the third line of this equation we have introduced the function $M(k)$ which maps the momentum 
$k$ in the DCA momentum cell to the cluster momentum contained in that cell.  Coarse-graining the first 
and the second terms on the right hand side of the above equation yields the cluster action (\ref{eqn:sclust}). 
Since the latter is independent of \tk, we may write
\begin{eqnarray}
S[c^{*},c] = \sum_{\tilde{k}}S_{\textrm{cluster}}[\overline{c}^{*},\overline{c}]
\;\;\;\;\;\;\;\;\;\;\;\;\;\;\;\;\;\;\;\;\;\;\;\;\;&&\nonumber\\
-\sum_{\omega,k,\sigma}c_{\omega,k,\sigma}^{*}[\Delta(M(k),i\omega)+\overline{h}_{M(k)}-h_{k}
]c_{\omega,k,\sigma}.&&
\label{eqn:srew}
\end{eqnarray}
Again, up to this point, we have only re-arranged terms and employed an identity which defines $\overline{c}$. 
 No approximation has been made.

The dual fermions are now introduced by means of the following Gaussian identity 
\begin{eqnarray}
\int \exp(-f^{*}_{\alpha} a_{\alpha\beta} f_{\beta} -f^{*}_{\alpha} b_{\alpha\beta} c_{\beta} 
- c^{*}_{\alpha} b_{\alpha\beta} f_{\beta}) \Pi_{\gamma} df^{*}_{\gamma} df_{\gamma} &&\nonumber\\
= \det (a) \exp[c^{*}_{\alpha} (b a^{-1}b)_{\alpha\beta} c_{\beta}]
\;\;\;\;\;\;\;\;\;\;\;\;\;\;\;\;\;\;\;\;\;\;\;\;\;\;\;\;\;\;\;\;&&
\label{eqn:identity}
\end{eqnarray}
for Grassmann variables in the path integral representation for the partition function
\begin{equation}
\int \exp(-S[c^{*},c]) \mathcal{D}[c^{*},c].
\end{equation}
To be specific, we choose the (diagonal) matrices according to
\begin{eqnarray}
a_{\omega,k,\sigma}&=&\bar{g}^{-2}(M(k),i\omega)[\Delta(M(k),i\omega)+\overline{h}_{M(k)}-h_{k}]^{-1};\nonumber\\
b_{\omega,k,\sigma}&=&\bar{g}^{-1}(M(k),i\omega).
\end{eqnarray}
where $\bar{g}$ is the single particle Green function calculated on the DCA cluster.
Applying the above identity to the second term in (\ref{eqn:srew}) yields
\begin{eqnarray}
%&&\frac{T}{N}\sum_{\omega,k,\sigma}f_{\omega,k,\sigma}^{*}g^{-2}(M(k),i\omega)[\Delta(M(k),i\omega)
%-h_{k}+\overline{h}_{M(k)}]^{-1}f_{\omega,k,\sigma}\nonumber\\
%&&+\frac{T}{N}\sum_{\omega,k,\sigma}f_{\omega,k,\sigma}^{*}g^{-1}(M(k),i\omega)c_{\omega,k,\sigma} 
%+ c_{\omega,k,\sigma}^{*}g^{-1}(M(k),i\omega)f_{\omega,k,\sigma}
&&\sum_{\omega,k,\sigma}\frac{f_{\omega,k,\sigma}^{*}\,f_{\omega,k,\sigma}}
{\bar{g}^{2}(M(k),i\omega)[\Delta(M(k),i\omega)+\overline{h}_{M(k)}-h_{k}]}\nonumber\\
&&+\sum_{\omega,k,\sigma}[f_{\omega,k,\sigma}^{*}\bar{g}^{-1}(M(k),i\omega)c_{\omega,k,\sigma} + h.c.].  
%%  There two terms are h.c to each other???  --shuxiang
\label{eqn:appltr}
\end{eqnarray}
The essential observation now is that, since $\bar{g}^{-1}(M(k),i\omega)\equiv \bar{g}^{-1}(K,i\omega)$ 
is independent of \tk, the second line of (\ref{eqn:appltr}) may be coarse-grained using
%\begin{equation}
%e^{A^{2}c_{\omega,k,\sigma}^{*}c_{\omega,k,\sigma}}=B^{-2}\int e^{-AB(c_{\omega,k,\sigma}^{*}
%f_{\omega,k,\sigma}+f_{\omega,k,\sigma}^{*}c_{\omega,k,\sigma}-B^{2}
%f_{\omega,k,\sigma}^{*}f_{\omega,k,\sigma}}df_{\omega,k,\sigma}^{*}df_{\omega,k,\sigma}\end{equation}
%where
%\begin{equation}
%A^{2}=\Delta(K,i\omega)-h_{k}+\overline{h}_{K}\end{equation}
%\begin{equation}
%B^{2}=\frac{1}{\overline{g}(K,i\omega)^{2}}\frac{1}{\Delta(K,i\omega)-h_{k}+\overline{h}_{K}}\end{equation}
%\begin{equation}
%\overline{g}(K,i\omega)=\frac{1}{i\omega+\mu-\overline{h}_{K}-\Delta(K,i\omega)}\end{equation}
again the DCA coarse-graining identity
\begin{equation}
\overline{f}_{\omega,K,\sigma}^{*}\overline{c}_{\omega,K,\sigma}\equiv\frac{N_{c}}{N}
\sum_{\tilde{k}}f_{\omega,K+\tilde{k},\sigma}^{*}c_{\omega,K+\tilde{k},\sigma}.\end{equation}
As a consequence the lattice action, Eq. (\ref{eqn:srew}), can be expressed in the form 
\begin{eqnarray}
S[c^{*},c;f^{*},f] = \sum_{\tilde{k}}S_{\textrm{restr}}[\overline{c}^{*},
\overline{c};\overline{f}^{*},\overline{f}]
\;\;\;\;\;\;\;\;\;\;\;\;&&\nonumber \\
%-\frac{T}{N}\sum_{\omega,K,\tk,\sigma}f_{\omega,K+\tk,\sigma}^{*}g^{-2}(K,i\omega)[\Delta(K,i\omega)
%+\overline{h}_{K}-h_{k}]^{-1}f_{\omega,K+\tk,\sigma}&&\nonumber\\
+\sum_{\omega,K,\tk,\sigma}\frac{f_{\omega,K+\tk,\sigma}^{*}\,f_{\omega,K
+\tk,\sigma}}{\bar{g}^{2}(K,i\omega)[\Delta(K,i\omega)+\overline{h}_{K}-h_{k}]}&&
\label{eqn:ccff}
\end{eqnarray}
where 
\begin{eqnarray}
S_{\textrm{restr}}[\overline{c}^{*},\overline{c};\overline{f}^{*},\overline{f}] 
= S_{\textrm{cluster}}[\overline{c}^{*},\overline{c}] 
\;\;\;\;\;\;\;\;\;\;\;\;\;\;\;\;\;\;\;\;\;\;\;\;\;\;&&\nonumber\\
+  \sum_{\omega,K,\sigma} 
[\overline{f}_{\omega,K,\sigma}^{*}\bar{g}^{-1}(K,i\omega)\overline{c}_{\omega,K,\sigma}+h.c.]&&
\end{eqnarray}
is the action \emph{restricted} to the cluster.

The transformation to dual fermions is completed 
by integrating out the fermionic degrees of freedom corresponding to 
$\overline{c}$ and $\overline{c}^{*}$. Since $S_{\textrm{restr}}$ is independent of \tk, 
this can be done individually for each cluster
\begin{eqnarray}
\frac{1}{Z_{\textrm{cluster}}}\int \exp(-S_{\textrm{restr}}[\overline{c}^{*},
\overline{c};\overline{f}^{*},\overline{f}])
\mathcal{D}[\overline{c}^{*},\overline{c}] 
\;\;\;\;\;\;\;\;\;\;\;\;\;\;\;\;&&\nonumber\\
= \exp\left(  -\sum_{\omega,K,\sigma} 
\overline{f}_{\omega,K,\sigma}^{*}\bar{g}^{-1}(K,i\omega)\overline{f}_{\omega,K,\sigma} 
- V[\overline{f}^{*},\overline{f}]\right). &&\nonumber\\
\label{eq:defineV}
\end{eqnarray}
Eq.~(\ref{eq:defineV}) defines the dual potential which can be obtained by expanding both sides 
and comparing the resulting expressions order by order. 
%To lowest order 
It is given by \cite{Hartmut_thesis}:
%
% An explicit expression for the dual potential is obtained by expanding both sides of 
%this equation and comparing the resulting expressions by order.
%The dual potential to lowest order reads
\begin{eqnarray}
V[\overline{f}^{*},\overline{f}] =\frac{1}{4}\sum_{KK'Q}\sum_{\omega\omega'\nu}
\sum_{\sigma_1,\sigma_2,\sigma_3,\sigma_4}
\;\;\;\;\;\;\;\;\;\;\;\;\;\;\;\;\;\;\;\;\;\;\;\;\;\;\;\;\;\;\;&&\nonumber\\
\gamma_{\sigma_1,\sigma_2,\sigma_3,\sigma_4}(K,K',Q;i\omega,i\omega',i\nu) 
\;\;\;\;\;\;\;\;\;\;\;\;\;\;\;\;\;\;&&\nonumber\\
\times\overline{f}_{\omega+\nu,K+Q,\sigma_1}^{*} \overline{f}_{\omega,K,\sigma_2}
\overline{f}_{\omega',K',\sigma_3}^{*}\overline{f}_{\omega'+\nu,K'+Q,\sigma_4}
\;\;\;\;&&\nonumber\\[3mm]
+\ldots\;\;\;\;\;\;\;\;\;\;\;\;\;\;\;\;\;\;\;\;\;\;\;\;\;\;\;\;\;\;\;\;\;\;\;\;\;\;\;\;\;\;\;\;\;\;\;\;\;\;\;\;\;\;\;\;\;\;&&
\label{eq:V_dual}
\end{eqnarray}
where $\gamma$ is the full (reducible) vertex of the cluster quantum impurity model, 
and the higher order contributions involve the $n$-body (for $n>2$) reducible vertices
as the bare interaction. Integrating out the lattice fermions results in an 
action which depends only on the dual fermion degrees of freedom given by
\begin{eqnarray}
S_d[f^{*},f] = -\sum_{k \omega \sigma}f^*_{\omega k \sigma} G^0_d(k,i\omega)^{-1}f_{\omega k \sigma} %\nonumber \\
 +\sum_{\tilde{k}} V[\overline{f}^*,\overline{f} ],
\label{eq:DFaction}
\end{eqnarray}
where $G^0_d$ is the bare dual Green function defined by
% The bare dual Green function is given by
\begin{equation}
%G_d^0(k,\omega) = - \overline{g}(K,\omega) \left[ \overline{g}(K,\omega) 
%+ \left(\Delta(K,i\omega) -h_{k}+\overline{h}_{K} \right)^{-1}  
%  \right]^{-1} \overline{g}(K,\omega)
G_d^0(k,i\omega) = - \frac{\overline{g}(K,i\omega)^2}{ \overline{g}(K,i\omega) 
+ \left(\Delta(K,i\omega)+\overline{h}_{K}  -h_{k}\right)^{-1}  }.
\label{Eq:dual-Green}
\end{equation}
This quantity together with the dual potential $V[\overline{f}^{*},\overline{f}]$
%, its n-body corrections (for $n>2$), and $G_d^0(k,i\omega)$ 
provides sufficient input for a many-body diagrammatic perturbation calculation on the dual lattice.

Note that besides the DCA coarse-graining process introduced here, the above derivation is 
a natural generalization of the dual fermion DMFA formulation of Rubtsov $et\,al.$ ~\cite{dual_fermion1}

% where the term $ \left( \Delta(K,i\omega) -h_{k}+\overline{h}_{K} \right) \sim {\cal{O}} (1/L_c^2) $ 
% where $L_c^2$ is the linear cluster size.  
% Following arguments which explain the rapid convergence of the DFDMF in the strong coupling limit, 
% The small nature of this term for large $L_c$ should ensure rapid convergence of the DFDCA.  In 
% particular, 
% \begin{eqnarray}
% G_d^0(k,\omega) &\approx& - \overline{g}(K,\omega) \left(\Delta(K,i\omega)
% -h_{k}+\overline{h}_{K} \right) \overline{g}(K,\omega) \nonumber\\
% &\sim& {\cal{O}}(1/L_c^2)
% \label{eq:G0dual}
% \end{eqnarray}
% with corrections of  ${\cal{O}}(1/L_c^4)$.  

\subsection{Self-consistency condition}

In rewriting the lattice action in terms of the cluster impurity model in the above derivation, 
the DCA hybridization function has been added and subtracted and hence is an arbitrary quantity. 
In order to fix this quantity we impose the condition
% that the coarse-grained bare dual Green function be zero
\begin{equation}
G_d^0(K,i\omega) = \frac{N_{c}}{N} \sum_{\tk} G_d^0(K+\tk,i\omega) \stackrel{!}{=}0.
\label{eqn:selfcons}
\end{equation}
To appreciate the consequences of this condition, first consider the DCA lattice Green function
\begin{equation}
G_{\textrm{DCA}}^{-1}(K+\tk,i\omega) = (i\omega+\mu)\mathbf{1}-h_{K+\tk} - \Sigma_{c}(K,i\omega),
\end{equation}
which can be expressed in terms of the cluster Green function
\begin{equation}
\bar{g}^{-1}(K,i\omega) = (i\omega+\mu)\mathbf{1}-\overline{h}_{K} - \Sigma_{c}(K,i\omega) - \Delta_{c}(K,i\omega),
\end{equation}
as
\begin{equation}
G_{\textrm{DCA}}^{-1}(K+\tk,i\omega) = \bar{g}^{-1}(K,i\omega) + \Delta_{c}(K,i\omega) +\overline{h}_{K} -h_{K+\tk}.
\end{equation}
Using the last expression, one may straightforwardly derive the following identity relating the 
DCA lattice Green function to the bare dual Green function
% [should we provide the derivation? 
%it is simple -3 lines, it depends a bit on the format of the final paper]:
\begin{equation}
G^{d,0}(K+\tk,i\omega) = G_{\textrm{DCA}}(K+\tk,i\omega) - \bar{g}(K,i\omega).
\end{equation}
Hence the above condition (\ref{eqn:selfcons}) is equivalent to requiring that the coarse-grained 
DCA lattice Green function be equal to the Green function of the cluster impurity model. 
This is exactly the DCA self-consistency condition. The DCA solution is therefore obtained 
if no diagrammatic corrections are taken into account and the hybridization is determined 
such that (\ref{eqn:selfcons}) holds. Consequently, we have a perturbation theory around the DCA 
as the starting point. While the DCA solution only depends on the cluster momentum K, 
the dependence on \tk\ can be introduced by solving the dual problem perturbatively.

\subsection{Scaling of the Dual Fermion DCA approach with cluster size}

The bare dual Green function is given by Eq.~(\ref{Eq:dual-Green}). 
%\begin{equation}
%G_{d}^{0}(k,\omega)=-\overline{g}(K,\omega)\left[\overline{g}(K,\omega)
%+\left(\Delta(K,i\omega)-h_{k}+\overline{h}_{K}\right)^{-1}\right]^{-1}\overline{g}(K,\omega)
%G_{d}^{0}(k,i\omega)=-\frac{\overline{g}(K,i\omega)^2}{\overline{g}(K,i\omega) +\left(\Delta(K,i\omega)+\overline{h}_{K}-h_{k}\right)^{-1}}.
%\end{equation}
If we introduce the  linear cluster size $L_c$ through $N_{c}=L_{c}^{D}$, one finds that
the term $\left(\Delta(K,i\omega)+\overline{h}_{K}-h_{k}\right)\sim\mathcal{O}(1/L_{c})$.
% and
%, in which 
%$L_{c}$ represents the
%Following arguments which explain the rapid convergence of the DFDMF in the strong coupling limit, 
The small nature of this term for large $L_{c}$ should ensure rapid convergence of the DFDCA. 
In particular, we then have
\begin{eqnarray}
G_{d}^{0}(k,i\omega)&=&-\overline{g}(K,i\omega)\left(\Delta(K,i\omega)+\overline{h}_{K}-h_{k}
\right)\overline{g}(K,i\omega)\nonumber\\
&&+\mathcal{O}(1/L_{c}^{2}),
\end{eqnarray}
i.e., the bare dual Green function also scales like
\begin{equation}
G_{d}^{0}(k,i\omega)\sim\mathcal{O}(1/L_{c}).
\label{eq:G0dual}\end{equation}
However, at points of high symmetry, where $\overline{h}_{K}-h_{k}  \sim\mathcal{O}(1/L_{c}^2)$, 
$G_{d}^{0}(k,i\omega)$ will fall more quickly than $\mathcal{O}(1/L_{c})$.  

\begin{figure}[tbh]
\includegraphics*[width=8cm]{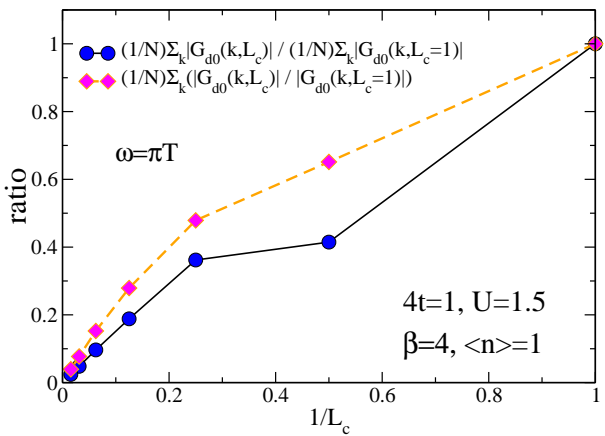}
\caption{Scaling plot for the bare dual Green function. Here we have used the 1-D Hubbard model 
to analyze its scaling behavior. Except for very small $L_c$ values, the two ratios scale linearly 
according to Eq.~(\ref{eq:G0dual}). } 
\label{fig:gd0_scaling}
\end{figure}

To illustrate the typical scaling behavior of the bare dual Green function, we plot in Fig.~\ref{fig:gd0_scaling}, as a function of $1/L_c$,
the ratio of $|G_{d}^{0}(k,i\omega=i\pi T, L_{c})|$ averaged over $k$ to the average 
$|G_{d}^{0}(k,i\omega=i\pi T, L_c=1)|$.  We also plot the average of the ratios.  The former initially falls more quickly 
than $\mathcal{O}(1/L_{c})$, while the latter displays a slower initial slope.  However, for $L_{c} \ge 4$ both fall 
roughly linearly in $1/L_{c}$. This behavior is found to be independent of temperature (not shown), since it is a purely 
algebraic effect.

Applying the standard tools to the dual fermion action, one obtains the formal expression
\begin{equation}
G_{d}(k,i\omega)=G_{d}^{0}(k,i\omega)+G_{d}^{0}(k,i\omega)T_{d}(k,i\omega)G_{d}^{0}(k,i\omega),\end{equation}
for the full dual fermion Green function $G_{d}(k,i\omega)$, where the reducible self-energy
 or scattering matrix $T_{d}(k,i\omega)$ of the dual system  is introduced. 
We will show later, that $T_{d}(k,i\omega)$ will be at most of order $\mathcal{O}(1/L_c^3)$, 
and we can infer the scaling
\begin{equation}
G_{d}(k,i\omega)\sim\mathcal{O}(1/L_{c})
\end{equation}
for the full dual fermion Green function too.

Once the dual fermion Green function is known, one can reconstruct the real lattice Green function as 
%can be shown to be
\begin{eqnarray}
G(k,i\omega)&=&\overline{g}(K,i\omega)^{-2}\left(\Delta(K,i\omega)+\overline{h}_{K}-h_{k}
\right)^{-2}G_{d}(k,i\omega)\nonumber\\
&&+\left(\Delta(K,i\omega)+\overline{h}_{K}-h_{k}\right)^{-1}\;.
\label{eqn:GrelGd}
\end{eqnarray}
Since $G(k,i\omega)$ is the Green function of the real lattice, it should scale as
\begin{equation}
G(k,i\omega)\sim\mathcal{\mathcal{O}}(1)
\end{equation}
with respect to any length scale.
On the other hand, 
%while 
for the two terms on the right hand side in (\ref{eqn:GrelGd}) we find
\begin{equation}
\overline{g}(K,i\omega)^{-2}\left(\Delta(K,i\omega)+\overline{h}_{K}-h_{k}
\right)^{-2}G_{d}(k,i\omega)\sim\mathcal{O}(L_{c})\end{equation}
and
\begin{equation}
\left(\Delta(K,i\omega)+\overline{h}_{K}-h_{k}\right)^{-1}\sim\mathcal{O}(L_{c})\;.
\end{equation}
Thus, the two $\mathcal{O}(L_{c})$ terms must cancel each
other. To verify this requirement, we insert the zeroth order contribution of the
dual Green function into the original Green function, and after some algebra we indeed obtain
\begin{eqnarray}
G(k,i\omega) \sim  \overline{g}(K,i\omega)\sim\mathcal{\mathcal{O}}(1),
\end{eqnarray}
with a correction given by
\begin{eqnarray}
\Delta G(k,i\omega) \sim  T_{d}(k,i\omega).
\end{eqnarray}
Therefore, the correction to the real Green function %correction calculated 
through the dual fermion approach scales the same way as the dual self-energy.

Presently, the dual potential Eq.~(\ref{eq:V_dual}) still contains 
an infinite hierarchy of vertices. The previous discussion now provides 
a very important insight into the contributions of these vertices to a perturbative expansion:
Each $n$-body diagrammatic insertion will involve a vertex and $n$ Green function lines.
In the parameter region away from a critical point the dual potential will be of order 
${\cal{O}}(1)$. As noted before, the dual Green function is of order ${\cal{O}}(1/L_c)$, i.e., 
each dual space diagrammatic insertion is of order ${\cal{O}}(1/L_c^2)$ 
when it involves the two-body dual space interaction, of order ${\cal{O}}(1/L_c^3)$ 
for the three-body interaction, and so on.  
%Terms of order $n$ in $V$ will contribute corrections of order ${\cal{O}}(1/L_c^{n})$. 
This means that the two-body contribution to $V$, explicitly shown in Eq.~(\ref{eq:V_dual}), 
will actually dominate and low-order perturbation theory will be sufficient to accurately capture the 
corrections to the DCA from the dual fermion lattice.

\subsection{Mapping back from the Dual-Fermion to the Real Lattice}

The relation of the real fermion Green function to the dual Green function 
has been been established in Eq.\ (\ref{eqn:GrelGd}). This is an exact relation 
which follows by taking the functional derivative of two equivalent partition functions.
They are linked through the same Gaussian identity that has been used to introduce 
the dual fermions (Eq. (\ref{eqn:identity})).
Higher order derivatives then allow us to derive relations between higher order cumulants. 
From this recipe, we find the following relation between the two-particle reducible vertex functions 
\begin{eqnarray}
F_{k,k',q;i\omega,i\omega',i\nu} & =& T(k+q,i\omega+i\nu) T(k,i\omega) \,F^{d}_{k,k',q;i\omega,i\omega',i\nu}\,\nonumber\\
&  \times & T(k',i\omega')T(k'+q,i\omega'+i\nu) %\nonumber\\
\end{eqnarray}
in real and dual space,
%\begin{eqnarray}
%&&\Gamma_{k,k',q;\omega,\omega',\Omega} \nonumber\\
%&=& T(k+q,\omega+\Omega) T(k,\omega) 
%\,\Gamma_{d}\, T(k',\omega')T(k'+q,\omega'+\Omega)\nonumber\\
%\end{eqnarray}
where
\begin{eqnarray}
T(k,i\omega) &=& \frac{G_{d}(k,i\omega)}{G(k,i\omega) (\Delta(K,i\omega)+\overline{h}_{K}-h_{k}) 
\bar{g}(K,i\omega)} \nonumber\\
&=& - [1+\bar{g}(K,i\omega)\Sigma_{d}(k,i\omega)]^{-1}.\label{eq:TKTP}
\end{eqnarray}
Similar relations hold for many-particle vertex functions.  With the help of 
the two-particle vertex function we can now express the corresponding susceptibility 
as $\chi=\chi_0+\chi_0 F \chi_0$. Since from Eq.\ (\ref{eq:TKTP}) it follows that $T(k,i\omega)$ is always 
finite, a divergence of $\chi$, signaling an instability or phase transition in real space, 
%in the dual space will also 
necessarily corresponds to an instability in the quantity $F^d$ in the dual fermion space. 
In order to locate the instabilities, it is hence sufficient to search for a divergence of the 
Bethe-Salpeter equation in the dual space.  For the special case when no diagrammatic corrections 
to the dual self-energy and vertex are taken into account, $T(k,i\omega) = -1$ and both DFDCA 
and DCA would produce the same phase diagram. In general cases, the DFDCA will produce results 
more realistic than DCA  due to the inclusion of additional long-ranged correlations 
from the dual fermion lattice diagrammatic calculation.

\section{Dual Fermion Diagrams}
In the DFDCA formalism, the dual fermion Green function is ${\cal{O}}(1/L_c) $(c.f.\ Eq.~(\ref{eq:G0dual})),
i.e., it acts as the small parameter in the diagrammatic expressions.  
In addition, in the strong coupling limit, the Green function is proportional to the hopping $t/U$,\cite{Hartmut_thesis} 
so each Green function leg contributes a factor of ${\cal{O}}((t/U)/L_c)$.  
In the weak coupling limit, the Green function remains ${\cal{O}}(1/L_c) $, 
but the vertices are now small,  with the two-body vertex behaving like ${\cal{O}}(U/t) $, 
the three-body vertex like ${\cal{O}}(U^2) $, and so on.  Each two-body diagrammatic insertion, 
composed of a two-body vertex and two dual fermion Green function legs, 
then scales like ${\cal{O}}(1/L_c^2) $, with an additional factor of $U$ or $t^2$ in the weak 
and strong coupling limits, respectively.  Each three-body diagrammatic insertion, 
composed of a three-body vertex and three dual fermion Green function legs, 
scales like ${\cal{O}}(1/L_c^3) $ with additional factors  $U^2$ or $t^3$ in the weak 
and strong coupling limits, respectively. 
%(Hartmut, is there a general rule for an the U-dependence of a n n-body dual fermion interaction? 
%- Yes, in general it is $U^{n-1}$ for an $n$-pariticle vertex, 
%so we have additional factors of $U^{n-1}$ and $t^n$ for an $n$-particle interaction).
\begin{figure}[tbh]
\includegraphics*[width=8cm]{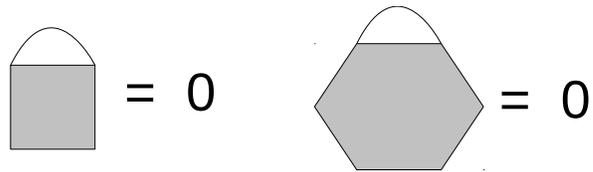}
\caption{Lowest order contributions to the dual fermion self-energy from the two-body interaction (left) 
and to the two-body interaction from the three-body term (right).  Since the bare n-body vertices 
depend only upon the small cluster $K$, the dual Green function line may be coarse-grained 
and are therefore zero according to Eq.~(\ref{eqn:selfcons}). } 
\label{fig:zerodiagrams}
\end{figure}

The boundary condition Eq.~(\ref{eqn:selfcons}) also constraints the diagrammatics.  
For example, the first-order contribution to the dual self-energy from the 2-body interaction 
is the Hartree-Fock contribution shown in Fig.~\ref{fig:zerodiagrams}.  
%\begin{equation}
%T_{d,2-body}^{1}\sim\gamma^{(4)}G_{d}=0\end{equation}
Since the vertex depends only upon the small cluster $K$, the dual Green function line 
may be coarse-grained.  The result is zero by virtue of Eq.~(\ref{eqn:selfcons}).  
Physically, 
this term must be zero since the Hartree term is already included 
in the cluster contribution to the self-energy.  Therefore, the first finite contribution 
to the dual fermion self-energy comes from the second order graph 
which contains three dual-fermion Green function lines.  This and all higher order contributions 
described by the Schwinger-Dyson equation already are of order $\mathcal{O}(1/L_{c}^{3})$, or smaller.  
Therefore, the fully dressed dual fermion Green function retains the scaling of the bare dual fermion Green function 
\begin{equation}
G_d(k,i\omega)  \sim {\cal{O}}(1/L_c)
\label{eq:Gdual}
\end{equation}
as anticipated earlier.

Similarly, the first-order 3-body contribution to the dual two-body vertex, 
also shown in \ Fig.~\ref{fig:zerodiagrams}, is zero. To see this, suppose the top leg is labeled 
by momentum $k = K + \tk$.  Since the remainder of the 3-body vertex does not depend upon $\tk$, 
we may freely sum over this label.  Again, the result is then zero through Eq.~(\ref{eqn:selfcons}).

%\begin{equation}
%T_{d,3-body}^{1}\sim\gamma^{(6)}G_{d}G_{d}\sim\mathcal{O}(1/L_{c}^{4})\end{equation}
As the cluster size becomes large, the DFDCA cluster problem may be accurately solved 
using low order perturbation theory,  keeping only the 2-body interaction vertex.  As described above,
two-body vertex insertion contributes an extra factor of $\mathcal{O}(1/L_{c}^{2})$, while
three-body vertex insertion contributes an extra factor of $\mathcal{O}(1/L_{c}^{3})$.  
It is therefore possible to use standard perturbation theory based on a two-body vertex 
to solve the dual-fermion DCA cluster problem, with an accuracy which turns out to be at least of 
order $\mathcal{O}(1/L_{c}^{4})$. 

For example, simple second order perturbation theory yields a self-energy $\mathcal{O}(1/L_{c}^{3})$.  
Two-body corrections, composed of a two-body vertex and two further Green function legs 
will contribute an extra factor $\mathcal{O}(1/L_{c}^{2})$ .  The first three-body contribution 
is the second order graph composed of one 2-body vertex and one 3-body vertex.  
It has four internal Green function legs, 
and is of order $\mathcal{O}(1/L_{c}^{4})$ so that the first three-body correction 
is smaller than the simple second order dual fermion self-energy composed of 2-body vertices 
by a factor of $\mathcal{O}(1/L_{c})$.  Self consistency, 
needed to impose the boundary condition Eq.~(\ref{eqn:selfcons}), is more important 
for the self-energy than higher order or three-body contributions.

\begin{figure}[tbh]
\includegraphics*[width=8cm]{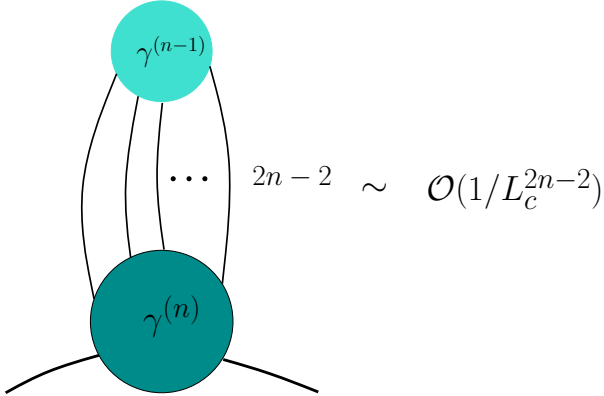}
\caption{Leading non-trivial n-body vertex contribution to the self-energy. It is contructed by 
one n-body vertex and one (n-1)-body vertex. Since there are (2n-2) internal legs, this 
contribution scales as $\mathcal{O}(1/L_c^{2n-2})$. } 
\label{fig:n-body}
\end{figure}

Generally, the leading non-trival n-body ($n\ge 3$) vertex contribution to the self-energy is constructed from
an $n$-body vertex and an $(n-1)$-body vertex, which are connected by $(2n-2)$ internal legs, 
as shown in Fig. \ref{fig:n-body}. Thus this contribution scales
as $\mathcal{O}(1/L_c^{2n-2})$.

\begin{figure}[tbh]
\includegraphics*[width=8cm]{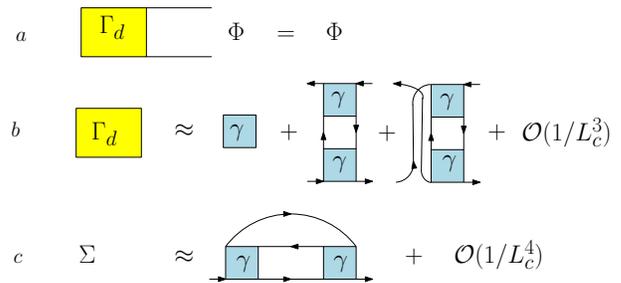}
\caption{(a) Equation for $T_c$. Transition temperatures on the
dual fermion lattice are identical to those calculated on the real lattice. 
(b) The low order corrections to the dual fermion irreducible vertex $\Gamma_d$.
The second order terms are of the order $\mathcal{O}(1/L_{c}^{2})$
with corrections $\mathcal{O}(1/L_{c}^{3})$.
(c) Contributions to the dual fermion self-energy. It is dominated by the second-order term
 which is of the order $\mathcal{O}(1/L_{c}^{3})$ with corrections $\mathcal{O}(1/L_{c}^{4})$. 
The self-energy adds relative corrections to $\chi_d^0$ of order $\mathcal{O}(1/L_{c}^{4})$
(see the text for the detail), so the most important contributions to the equation for $T_c$
come from the second-order cross channel contributions to $\Gamma_d$.} 
\label{fig:Tcequation}
\end{figure}

As another example, consider the equation for a transition, 
in the pairing matrix formalism (Fig.~\ref{fig:Tcequation})
\begin{equation}
\Gamma_d \chi_d^0 \Phi = \Phi
\end{equation}
where $\Phi$ is the leading eigenvector of the pairing matrix $\Gamma_d \chi_d^0$. 
 A transition is indicated by the corresponding eigenvalue approaching one. 
To lowest order, the irreducible dual fermion vertex $\Gamma_d\approx \gamma$ is just the bare dual fermion interaction, 
and the legs in $\chi_d^0$ are not dressed by the dual fermion self-energy.  
In this case the transition temperatures of the DCA are reproduced (e.g., see Fig.~\ref{fig:sigma_d0}).  
The lowest order corrections to the DCA come from the second order corrections to the vertex, 
which contain two dual fermion Green function legs and are therefore $\mathcal{O}(1/L_{c}^{2})$.  
The low order contributions to $\chi_d^0$ are $\chi_d^0 \approx  G_d^{0}(1 +  \Sigma_d G_d^{0} +\cdots )G_d^{0}$,
and thus the lowest relative correction to $\chi_d^0$ is of order $\mathcal{O}(1/L_{c}^{4})$.  
Therefore, the cross channel second order corrections 
to the vertex are more important than the second-order corrections to the self-energy 
when the DCA cluster size is large.  We note that this is not only true for the DFDCA, 
but also for the DFDMFA in the strong coupling limit where the small parameter $t/U$ 
replaces $1/L_{c}$.  Furthermore, higher order approximations such as the ladder approximation
 that do not include these cross channel contributions are not appropriate 
for the solution of the dual fermion lattice in the limit of large DCA cluster size or small $t/U$.

Higher order approximations like the fluctuation-exchange approximation (FLEX)~\cite{FLEX}, 
which include the cross channel contributions to $\Gamma_d$, should on the other hand be quite accurate.  
In fact, the FLEX contains all two-particle diagrams to third order.  The first diagram 
neglected by the FLEX is composed of one three-body and one two-body vertex and would contribute a 
correction $\mathcal{O}(1/L_{c}^{4})$ to the self-energy or $\mathcal{O}(1/L_{c}^{3})$ to the vertices.  

\section{Results} 
In this section, we will present numerical results from a DFDCA calculation, where the interaction expansion 
continous-time quantum Monte Carlo method\cite{ctqmc} is employed to solve the cluster problem within the DCA calculation.
We will restrict the discussion to the two-dimensional Hubbard model on the square lattice 
with only nearest neighbor hopping. Thus, for half-filling we expect 
strong antiferromagnetic correlations, which will drive an antiferromagnetic transition within DCA.
In this case, as the Mermin-Wagner theorem prohibits long-range order except at zero temperature,
we expect strong renormalization of the N\'eel temperature, $T_N$, from DFDCA.

\begin{figure}[t]
\includegraphics*[width=8cm]{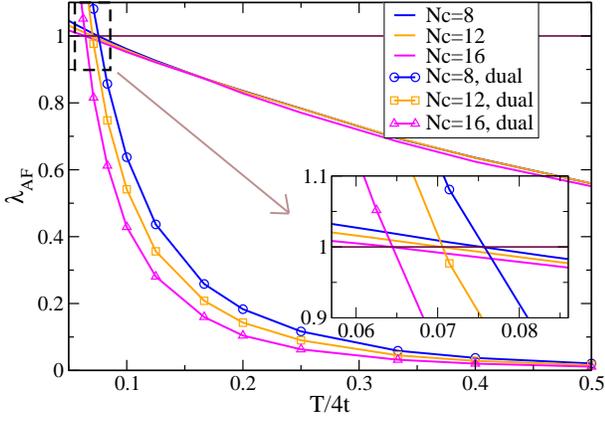}
\caption{(Color online) Plots of leading eigenvalues for different cluster sizes 
for the anti-ferromagnetic channel with $U=6t$ and filling 
$\langle n\rangle=0.95$. 
Lines without symbols are results from the DCA calculation for clusters with sizes
$N_c=8$, $12$ and $16$, while lines with symbols are results 
from the DFDCA calculation without self-energy correction.
For the latter, we have used a linear size of the dual fermion lattice as large as
several hundreds ($N=L\times L$, $L\sim 200$).
The inset is an enlarged view around the transition point. 
Note that both calculations produce the same transition temperatures as expected.} 
\label{fig:sigma_d0}
\end{figure}

To check the correctness of our implementation of the DFDCA approach, we first carry out 
calculations with the correction from the dual-fermion lattice turned off. 
For this trivial case, one expects DFDCA to reproduce the same physics as  DCA. 
Fig.~\ref{fig:sigma_d0} displays the leading eigenvalues for different cluster sizes 
at filling $\langle n \rangle = 0.95$ for the antiferromagnetic channel. 
Note that for each cluster size, both the DFDCA and the DCA leading eigenvalues
cross the line $\lambda=1$ at the same temperature, which is the mean-field N\'eel temperature, 
and that with increasing cluster size $T_N$ decreases, as expected.
It is also interesting to note that the DFDCA provides a sensitive way to monitor 
the finite-temperature transitions since the DFDCA leading eigenvalues have a 
steeper slope when crossing the $\lambda=1$ line.

For the non-trivial DFDCA calculation, we expect to see for a fixed cluster size 
a reduction of the N\'eel temperature since  correlations 
beyond the cluster scale are now incorporated by
\begin{figure}[t]
\includegraphics*[width=8cm]{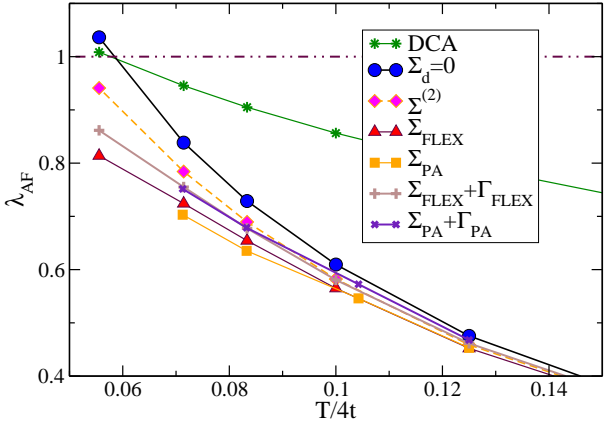}
\caption{(Color online) Plots of leading eigenvalues for the anti-ferromagnetic channel.
They are calculated with different approximate methods in the dual-fermion lattice.
The parameters used are $U=4t$, $\langle n\rangle=1$, the DCA cluster of size $N_c=1$, and 
the dual fermion lattice of a size $N=4\times 4$.}
\label{fig:lev_Nc1}
\end{figure}
the dual fermion calculation. For the dual fermion lattice, 
we employ different approximation schemes:  the self-consistent second-order 
perturbation theory (SOPT), FLEX and the parquet approximation (PA) \cite{parquet}. 
The results are collected in Fig.~\ref{fig:lev_Nc1}, 
where the power of DFDCA manifests itself clearly. The simple second-order correction 
from the self-energy is already able to reduce the N\'eel temperature
by ten percent. Taking into account more Feynman diagrams with higher orders, 
for example by FLEX or PA,  continues to reduce the N\'eel temperature. 
However, the inclusion of vertex correction tends to increase the N\'eel temperature again. 
For example, the eigenvalues labeled $\Sigma_{\rm{FLEX}}$ are calculated with a bare 
dual fermion vertex and FLEX dressed legs, while those labeled $\Sigma_{\rm{FLEX}}+\Gamma_{\rm{FLEX}}$
are calculated with both FLEX dressed legs and vertex (see b and c in Fig. \ref{fig:Tcequation} 
for contributions up to second-order in the bare dual fermion vertex $\gamma$). %{\color{red}Any explanation???}

\begin{figure}[t]
\includegraphics*[width=8cm]{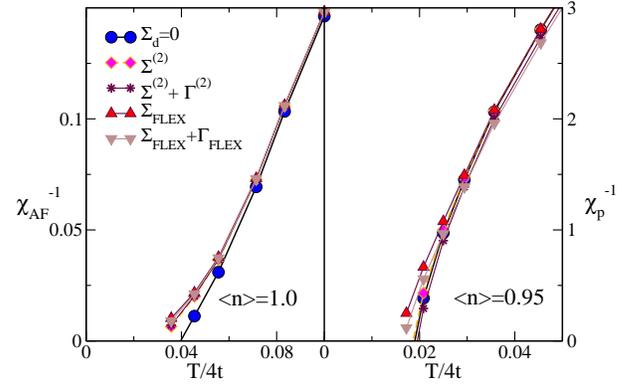}
\caption{(Color online) Plots of the inverse anti-ferromagnetic and d-wave pairing susceptibilities 
calculated with different approximate methods in the dual fermion calculation. 
The parameters used are $U=8t$ and $N_c=4$. The linear dependence of the results with $1/L^2$ 
(see Fig.~\ref{fig:lev_Lc2}) is used to extrapolate the $L=\infty$ limit results.}
\label{fig:chi_Nc4}
\end{figure}

Up to now we have only discussed the leading eigenvalues of the vertex function. 
Of course, the DFDCA also allows to calculate the full susceptibility 
from the Bethe-Salpeter equation. Two typical results for an $N_c=2\times 2$ 
DCA cluster are shown in Fig.~\ref{fig:chi_Nc4}, as function of temperature 
for $U=8t$. In the left panel, the inverse staggered susceptibility 
for half filling is displayed, while the right one contains results 
for the inverse $d$-wave pairing susceptibility at a filling $\langle n\rangle =0.95$. 
Due to the heavy computational cost for the parquet calculation, here we only use the SOPT and FLEX in our dual fermion lattice calculation.
Although $L_c=2$ is not really large, 
%the short length scale correlations are included in DCA when $N_c>1$, 
the DFDCA is still able to significantly reduce the mean-field N\'eel 
and the abnormally large superconducting transition temperatures.
It is quite interesting to note, that for the anti-ferromagnetic channel at half-filling, 
SOPT and FLEX produce similar results,  both being different from 
the DCA results. The effect of vertex correction is small in this case. 
For the $d$-wave pairing susceptibility at $\langle n\rangle=0.95$, on the other hand,
SOPT in dual space makes almost no difference from the DCA results, but
the FLEX tends to significantly reduce the pairing susceptibility. Again, the inclusion of vertex correction
has the opposite effect, i.e., it leads to a slight increase of the critical temperatures.

\begin{figure}[t]
\includegraphics*[width=8cm]{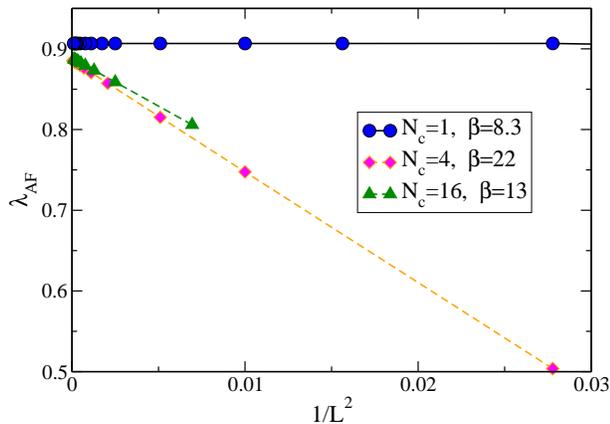}
\caption{(Color online) The $L$ dependence of the leading eigenvalue for different DCA clusters.
The parameters used are $U=8t$, $\langle n\rangle=1$, $\Sigma_{dual}=0$ and $\beta=4t/T$.
The nice linear dependence of $1/L^2$ can be readily observed, which is due to the periodic boundary conditions
used in the dual-fermion calculation. 
}
\label{fig:lev_Lc2}
\end{figure}

In the derivation of the DFDCA approach, we have assumed that the dual fermion lattice is infinite. 
However, in practical calculations, the size is limited due to the algebraic increase of 
the computational cost. This results in some deviations from the infinite size system. Fig.~\ref{fig:lev_Lc2} shows
the $L$ ($N=L\times L$) dependence of the leading eigenvalues for different DCA clusters.
The nice linear dependence of the leading eigenvalues on $1/L^2$ can be readily observed. 
This is due to the periodic boundary conditions used in the dual-fermion calculation. 
This property allows us to reduce the computational cost of our calculation 
by using two small $L's$ and extrapolating to obtain a rather accurate approximation of the $L=\infty$ result.

\section{Discussion}

The dual fermion mapping as discussed in section~\ref{sec:DFDCA_derivation} 
is exact, and the approximation is made only when performing the diagrammatic calculation 
for the dual fermion lattice. Justified by the scaling behavior of the dual fermion Green function, 
it suffices to consider the 2-body term of the interaction and use low order perturbation theory.
Correlations beyond the DCA cluster size are systematically restored through the dual fermion 
calculation on the lattice.   % I think it's better to put this sentence back here. --shuxiang
In this sense, the DFDCA can be seen as a diagrammatic expansion 
around DCA. This is manifested in Fig.~\ref{fig:sigma_d0} and \ref{fig:lev_Nc1} 
where we see that not including the dual fermion self-energy and vertex corrections 
reproduces the DCA transition temperature. When these corrections are included, 
we observe a systematic suppression of the DCA transition temperature resulting 
in a more realistic value. This is clearly seen 
in Fig.~\ref{fig:lev_Nc1} and \ref{fig:chi_Nc4}.
Since correlations at intermediate length-scale are taken into account by 
the dual fermion lattice calculation, we can use small clusters in the underlying DCA calculation.
As a result, we are able to greatly reduce the adverse effect of the minus sign problem encountered
in QMC simulations for larger clusters, and access a wider region of parameter space.  

The DFDCA has an additional advantage that it is parameterized by 
the full (reducible) vertex function calculated on the DCA cluster.  Other multi-scale
methods\cite{DVA, MSMB, Kusunose} rely upon the calculation of the cluster irreducible or fully irreducible 
vertices.  Our recent numerical experiments show that inverting the Bethe-Salpeter equation to obtain
the irreducible vertex, which is also the first step in the calculation of the fully irreducible
vertex,  fails in some parameter regions, especially for large $U$ or near half
filling. This difficulty is avoided in the DFDCA. 

The dual fermion mapping also assumes that the dual fermions are treated on an infinitely large lattice. 
In practice however, they are treated on a finite-size lattice. 
Thanks to the finite-size scaling behavior observed in Fig.~\ref{fig:lev_Lc2},
finite-size calculations are used to extrapolate to the infinite-size lattice, 
leading to a reduction of the computational cost in the dual fermion lattice calculation.

Note that in the calculations presented here, we have not performed the 
full self-consistency where the dual fermion result is used to determine 
the DCA cluster hybridization that is fed back into the DCA calculation until convergence. 
However, this first iteration already produces more satisfactory values 
for the N\'eel temperature as well as the d-wave superconducting transition temperature. 
We can anticipate that the full self-consistency will further improve the performance of this approach.

With the full self-consistency implemented and tested, we are planning to apply this approach to map out 
the phase diagram for the 2-D Hubbard model in the hole-doped region. In addition, 
we are also working on applying this approach to the Falicov-Kimball model and the Anderson disorder model recently.

%Note that in our numerical calculations, we have not fed the results calculated from the dual fermion lattice 
%back to the DCA calculation. However, this first iteration is already able to improve the results by
%reducing the mean field N\'eel temperature and d-wave superconducting temperature to more realistic ones. And
%we would expect to have more improvement by doing the calculation in a more self-consistent manner, which
%will be our next step work. 

%And about the dual fermion lattice calculation, we have used the finite size scaling behavior to extrapolate
%the infinite size lattice results. One can also employ DCA for this finite size calculation and would expect
%to reduce the computational cost further. 

\section{Conclusion} 
We have designed a new multi-scale many body approach, the Dual Fermion Dynamical Cluster approach (DFDCA), 
by combining the DCA and the recently 
introduced dual fermion formalism. The DFDCA uses both single and two particle quantities calculated
in DCA as the input for the dual fermion calculation. Different self-consistent diagrammatic approximations can be used in the dual 
fermion lattice, which systematically restores the long-ranged correlation ignored during 
the DCA calculation.

This approach is a systematic expansion around the DCA calculation. 
Our numerical experiments show that the zeroth order result ($\Sigma_d=0$) reproduces 
the original DCA, and for any non-trivial dual fermion calculation, it is an improvement on the 
DCA calculation. We applied different self-consistent diagrammatic methods, self-consistent 2nd-order, 
FLEX and parquet approximation, on the dual fermion lattice. 
They all improved the DCA calculation by reducing the mean-field N\'eel temperature by different amounts.
In addition, the abnormally large superconducting transition temperature of the four site cluster calculation
can be reduced by this approach as well.

%\section{Acknowledgments}   
\begin{acknowledgments}
 
We would like to thank A. I. Lichtenstein for useful conversations. This research is supported by DOE 
SciDAC DE-FC02-06ER25792 (SXY, HF, KMT, and MJ)  and  NSF grants OISE-0952300 (SXY, HF, and JM).  
This research was supported in part by NSF through TeraGrid resources provided by 
the National Institute for Computational Sciences under grant number TG-DMR100007 and by the 
high performance computational resources provided by Louisiana State University (http://www.hpc.lsu.edu).

\end{acknowledgments}


\begin{thebibliography}{99}
\bibitem{explanation} Note that since the interaction is assumed to be local, it is unaffected 
by coarse-graining. Non-local interactions however will be coarse-grained.
\end{thebibliography}
\end{document}